\documentclass{article}
\usepackage{spconf,amsmath,graphicx}
\usepackage[ruled,vlined]{algorithm2e}
\usepackage{url}

\usepackage{amsmath,amsfonts,bm, bbm}

\def\eqref#1{equation~\ref{#1}}

\def\1{\bm{1}}

\def\rvb{{\mathbf{b}}}

\def\rmU{{\mathbf{U}}}
\def\rmV{{\mathbf{V}}}
\def\rmW{{\mathbf{W}}}

\def\vdelta{{\bm{\delta}}}

\def\vb{{\bm{b}}}

\def\vg{{\bm{g}}}
\def\vh{{\bm{h}}}

\def\vx{{\bm{x}}}
\def\vy{{\bm{y}}}

\def\mI{{\bm{I}}}

\def\mU{{\bm{U}}}

\def\mW{{\bm{W}}}

\DeclareMathAlphabet{\mathsfit}{\encodingdefault}{\sfdefault}{m}{sl}
\SetMathAlphabet{\mathsfit}{bold}{\encodingdefault}{\sfdefault}{bx}{n}

\def\E{{\mathbb{E}}}

\usepackage{mdframed}

\usepackage{accents}
\newlength{\dhatheight}

\newcommand\blfootnote[1]{%
  \begingroup
  \renewcommand\thefootnote{}\footnote{#1}%
  \addtocounter{footnote}{-1}%
  \endgroup
}

\title{Wearing a MASK: Compressed Representations of Variable-Length \\ Sequences
Using Recurrent Neural Tangent Kernels}
\name{Sina Alemohammad, Hossein Babaei, Randall Balestriero, Matt Y.\ Cheung, Ahmed Imtiaz Humayun}{
Daniel LeJeune, Naiming Liu, Lorenzo Luzi, Jasper Tan, Zichao Wang, Richard G.\ Baraniuk}
\address{Rice University}
\begin{document}
%
\maketitle
\begin{abstract}
High dimensionality poses many challenges to the use of data, from visualization and interpretation, to prediction and storage for historical preservation. Techniques abound to reduce the dimensionality of fixed-length sequences, yet these methods rarely generalize to variable-length sequences. To address this gap, we extend existing methods that rely on the use of kernels to variable-length sequences via use of the Recurrent Neural Tangent Kernel (RNTK). Since a deep neural network with ReLu activation is a Max-Affine Spline Operator (MASO), we dub our approach Max-Affine Spline Kernel (MASK). We demonstrate how MASK can be used to extend principal components analysis (PCA) and 
t-distributed stochastic neighbor embedding (t-SNE) and apply these new algorithms to separate synthetic time series data sampled from second-order differential equations. \blfootnote{This work was supported by NSF grants CCF-1911094, IIS-1838177, IIS-1730574, and 1842494; ONR grants N00014-18-12571 and N00014-20-1-2787; AFOSR grant FA9550-18-1-0478; and a Vannevar Bush Faculty Fellowship, ONR grant N00014-18-1-2047. \nonumber
}
\end{abstract}
\begin{keywords}
Dimensionality reduction, Variable-length, Neural tangent kernel, t-SNE, Recurrent neural network 
\end{keywords}

\section{Introduction}
\label{sec:intro}

Many signal processing and machine learning problems involve high-dimensional data.
In order to analyze and exploit such data effectively, we must find meaningful low-dimensional representations. On the one hand, it is often extremely valuable to visualize data as a part of the exploration process, yet we are limited as humans to visualizing in only a few dimensions, typically two or three. On the other hand, the data may be nonlinearly embedded in a high-dimensional space, yet the meaningful attributes for a prediction task of interest may lie in a low-dimensional space where linear predictors achieve high accuracy. 

Consequently, myriad methods for dimensionality reduction have been developed. These include linear parametric transformations, such as Principal Components Analysis (PCA) \cite{pearson1901liii}, sparse PCA \cite{zou2006sparse}, and Independent Components Analysis (ICA) \cite{comon1994independent}; nonlinear parametric methods, such as bottleneck features \cite{grezl2007probabilistic}; and non-parametric methods, such as multi-dimensional scaling (MDS) \cite{cox2008multidimensional}, Isomap \cite{tenenbaum2000global} and t-distributed stochastic neighbor embedding (t-SNE), all of which transform data points from the input space to a low-dimensional space. Many of these methods find embeddings for collections of data points in new spaces that preserve similarities or distances. For example, in classical MDS, data points are embedded into a space of lower dimensions to preserve pairwise Euclidean distance. Such a distance, while simple and efficient, yields a linear embedding that is incapable of preserving nonlinear structures in data, corroborating research for more representative distance measures that replace Euclidean distance.
Isomap preserves the geodesic distance of the points \cite{tenenbaum2000global}, and t-SNE preserves distances based on fitted probability distributions \cite{maaten2008visualizing}. Some works take these a step further and use pairwise distances based on kernel functions, a technique commonly used in machine learning. Such kernel functions have successfully been applied to PCA \cite{mika1999kernel}, Isomap \cite{choi2004kernel, choi2007robust} and t-SNE \cite{gisbrecht2015parametric}.

Unfortunately, a challenge to all these methods is that the pairwise distance function requires the two data points under comparison to have the same number of dimensions, restricting these methods to fixed-length data. This is particularly troubling for time-series data such as natural language or physics continuity equations~\cite{sst2013,speech-recognition,bird}, wherein one may need to deal with sequences of different lengths. One way to solve this is to use a pairwise distance function that can take inputs of different lengths.

Recent advances in deep learning theory have revealed that the training dynamics of infinite-width neural networks (NNs) learned via first-order gradient descent with Gaussian parameter initialization are captured by a so-called {\em neural tangent kernel} (NTK) \cite{jacot2018neural}. In particular,~\cite{alemohammad2020recurrent,yang2020tensor} showed that a kernel can be derived from an infinitely wide recurrent neural networks called the recurrent neural tangent kernel (RNTK).
The RNTK inherits the universal representation power of infinitely wide RNNs~\cite{10.1007/11840817_66} while also representing a practical kernel that has been shown to outperform both kernels induced from non-recurrent neural architectures and classical kernels on a number of time series tasks~\cite{alemohammad2020recurrent}.  
RNTK also gracefully handles data-points of varying lengths.
The above results and properties of RNTK make it a promising yet under-explored candidate for time-series dimensionality reduction applications. 

\subsection{Contributions}
We propose {\it MASK} (Max-Affine Spline Kernel), a kernel-based time series dimensionality reduction method based on RNTK with ReLU activation (and thus can be regarded as a max-affine spline operator (MASO); see~\cite{wang2018a,pmlr-v80-balestriero18b} for detailed explanations). 
We show that by simply replacing the commonly used kernels (such as Gaussian and polynomial) with the RNTK in classic dimensionality reduction algorithms such as PCA and and t-SNE, we can enable these algorithms 
(which are traditionally not well suited for time series data) 
to work well with time series data.
We experimentally validate our proposed MASK with PCA (RNTK PCA) and t-SNE (RNTK t-SNE) on several realistic synthetic time series data generated by linear dynamical systems and demonstrate MASK's superior performance in discovering clear patterns in each dataset in the reduced dimensional space compared to PCA and t-SNE with classic kernels.


\section{RNTK as a Similarity Measure}
In this section, we introduce RNTK and briefly show how to compute it given a set of time series sequences of potentially varying lengths. 
Note that this kernel matrix can be treated as a similarity measure which then leads to its application to dimensionality reduction that we explain in more detail in the next section.

Consider a Deep Neural Network (DNN) mapping $f_{\theta}(\vx)$ with parameters $\theta$ and a pair of inputs $\vx$ and $\vx'$, the {\em Neural Tangent Kernel} (NTK) \cite{jacot2018neural} is defined as:
\begin{align}
\label{eq:ntk}
    \Theta(\vx,\vx') := \langle \nabla_{\theta}f_{\theta}(\vx) , \nabla_{\theta}f_{\theta}(\vx')\rangle.
\end{align}
where $\theta$ is Gaussian distributed.  In the regime of infinite width of the layers employed by $f_{\theta}$, the above converges to an analytical form \cite{lee2017deep, duvenaud2014avoiding,novak2018bayesian, garrigaalonso2018deep,yang2019tensor} which can be used to compute input similarities or gain insights into DNNs training and representation power \cite{du2018gradient, NIPS2019_8893, allen2019learning, zou2018stochastic}.
When the consider DNN as a recurrent architecture, namely a Recurrent Neural Network, the limiting kernel is denoted as the  RNTK \cite{alemohammad2020recurrent, yang2020tensor}. The key property that we will exploit in this paper is the fact that RNTK provides a novel way to obtain similarity measure of different length inputs thanks to the underlying recurrent modeling of the inputs.

We overview the derivation of the RNTK from a single-layer RNN with average pooling \cite{ElmanRNN}. Given an input data sequence $\vx = \{ \vx{(t)} \}^{T}_{t = 1}$ of fixed length $T$ with $\vx^{(t)} \in \mathbb{R}^m$, an RNN with $n$ units in each hidden layer performs the following recursive computation at each layer and each time step 
\begin{align} 
	\vg^{\left(t\right)}(\vx) &= \frac{\sigma_w}{\sqrt{n}}\rmW \vh^{\left(t-1\right)}(\vx) +  \frac{\sigma_u}{\sqrt{n}}\rmU \vx{(t)} + \sigma_b\rvb  \\
	\vh^{\left(t\right)}(\vx) &= \phi\left(  \vg^{\left(t\right)}(\vx) \right)\, 
\end{align}
where $\mW \in \mathbb{R}^{n \times n}$, $ \vb \in \mathbb{R}^{n}$, $ \mU \in \mathbb{R}^{n \times m} $, and $\phi(\cdot) : \mathbb{R} \rightarrow \mathbb{R}$ is the activation function that acts entry-wise on a vector. 	$\vh^{\left(0\right)}(\vx)$ is the initial hidden state, which is set to zero. The output of an RNN at time $t$ is computed via
\begin{align}
     f^{(t)}_\theta(\vx) = \frac{1}{\sqrt{n}}\rmV \vh^{(t)}(\vx)
\end{align}
and the final output of the RNN is computed via
\begin{align} 
    f_\theta(\vx) &= \sum_{t = 1}^{T} f^{(t)}_\theta(\vx) =  \sum_{t = 1}^{T} \frac{1}{\sqrt{n}}\rmV \vh^{(t)}(\vx) \in \mathbb{R}^d. 
\end{align} 
The learnable parameters $\theta := \mathrm{vect}\big[\rmW,\rmU,\rvb , \rmV \} \big]$ are initialized to Gaussian $\mathcal{N}(0,1)$ random variables. The set $\mathcal{S} = \{\sigma_w,\sigma_u,\sigma_b \}$ represents the initialization hyperparametrs.

Leveraging the equivalence of infinite-width NNs with Gaussian processes (GPs) \cite{lee2017deep, duvenaud2014avoiding,novak2018bayesian, garrigaalonso2018deep,yang2019tensor}, as $n \rightarrow \infty$, each coordinate of the RNN output at each time step converges to a Gaussian process for two input sequence of possible different length $\vx = \{ \vx{(t)} \}^{T}_{t = 1}$ and $\vx' = \{ \vx'{(t)} \}^{T'}_{t = 1}$ with kernel
\begin{align} 
    \mathcal{K}^{(t,t')}(\vx,\vx') & =  \mathop{\E}_{\theta \sim \mathcal{N}}\big[ [f^{(t)}_{\theta}(\vx)]_i \boldsymbol{\cdot} [f^{(t')}_{\theta}(\vx')]_i 
    \big], \hspace{1mm} \forall i \in [d]
\end{align}
known as the NN-GP kernel. The pre-activation $\vg^{\left(t\right)}(\vx)$ and gradient vectors $\vdelta^{(t)}(\vx) := \sqrt{n} \big( \nabla_{\vg^{(t)}(\vx)} f_{\theta}(\vx) \big)$ also converge to zero mean GPs with kernels 
\begin{align} 
    \Sigma^{(t,t')}(\vx,\vx') &= \mathop{\E}_{\theta \sim \mathcal{N}} \big[ [\vg^{(t)}(\vx)]_i \boldsymbol{\cdot} [\vg^{(t')}{(\vx')}]_i  \big]\, \,\, \forall i \in [n] 
    \\
    \Pi^{(t,t')}(\vx,\vx') &= \mathop{\E}_{\theta \sim \mathcal{N}} \big[ [\vdelta^{(t)}(\vx)]_i \boldsymbol{\cdot} [\vdelta^{(t')}(\vx')]_i  \big]\, \,\, \forall i \in [n].
\end{align} 
The RNTK for two inputs $\vx$ and $\vx'$ with length $T$ and $T'$ can thus be obtained as
\begin{align}
    \Theta(\vx,\vx') = &\Bigg( \sum_{t =  1}^{T} \sum_{t' = 1}^{T'} \left( \Pi^{(t,t')}(\vx,\vx') \boldsymbol{\cdot} \Sigma^{(t,t')}(\vx,\vx') \right) \nonumber\\
    &\quad + \mathcal{K}^{(t,t')}(\vx,\vx') \Bigg)  \otimes \mI_d. \,
\end{align}
where $\mI_d$ is the identity matrix of size $d$. For $\phi(z) = \mathrm{ReLU}(z) = \rm max (0,z)$, an analytical formula exists for the RNTK, enabling fast point-wise evaluation of RNTK on the input data. The full and detailed derivation of the RNTK is given in \cite{yang2020tensor}. Given a means to compute the RNTK, we can use the RNTK to compute similarities between data points of varying length and extend classical dimensionality reduction techniques.

\section{MASK: RNTK-Based Dimensionality Reduction}

We now present MASK, the incorporation of the RNTK into standard dimensionality reduction methods. The key idea is to replace the pairwise distance (or similarity) measure used in these methods with a new distance computed by RNTK. We introduce two specific instances of MASK: RNTK PCA and RNTK t-SNE.
While we specifically use PCA and t-SNE as examples, we emphasize that MASK is generally applicable and that any dimensionality reduction technique that operates off of kernels is compatible with MASK.

\subsection{RNTK PCA}

PCA constructs new dimensions by finding the linear combination of the original dimensions that are uncorrelated with each additional dimension containing the most variance possible from the data \cite{abdi2010principal}. It can be shown that these dimensions are the eigenvectors of the data's covariance matrix and are thus linear combinations of the original data dimensions, yielding a linear embedding. Many times, the data is not linearly separable in its original dimensions, and it may be beneficial to first map the data points into a higher-dimensional space before finding its covariance matrix's eigenvectors to allow nonlinear embeddings. Using the kernel trick, one need not compute the higher-dimensional data points; one simply needs to compute the inner products between the higher-dimensional data points (i.e.\ the output of the kernel function) in the form of a matrix $K$ with entries $K_{i, j} = \langle \Phi(\vx_i), \Phi(\vx_j) \rangle$ for some given mapping function $\Phi$ and following the method in \cite{scholkopf1997kernel}. RTNK PCA is obtained by using the RTNK outputs as the kernel: $K_{i, j} = \Theta(\vx_i, \vx_j)$ from Eq.\ \ref{eq:ntk}.

\subsection{RNTK t-SNE}

In t-SNE, the pairwise similarity of two data points $\vx_i$ and $\vx_j$ is defined as a conditional probability:
\begin{align}
   p_{j|i} = \frac{\exp\left(\tfrac{-\|\vx_i, \vx_j\|^2}{2\sigma_i^2}\right)}{\sum_{k\neq i}\exp\left(\tfrac{-\|\vx_i-\vx_j\|^2}{2\sigma_i^2}\right)},
\end{align}
where $\sigma_i$ is a parameter \cite{maaten2008visualizing}. Next, a probability distribution is defined based on the distances of the low-dimensional embeddings:
\begin{align}
    q_{j|i} = \frac{\exp(-\|\vy_i - \vy_j\|^2)}{\sum_{k\neq i}\exp(-\|\vy_i - \vy_k\|^2)}.
\end{align}
The low-dimensional embeddings $\vy_i$ for each data point $\vx_i$ is then obtained by minimizing the Kullback-Leibler divergence $C = \sum_i \sum_j p_{j|i}\log\frac{p_{j|i}}{q_{j|i}}$ between the two induced probability distributions. 

To apply the RNTK function into t-SNE, we first define the RNTK distance based on Eq.\ \ref{eq:ntk} and the Pearson dissimilarity \cite{solo2019pearson}:
\begin{align}
\label{eq:rntk_dist}
d(\vx_i, \vx_j) 
= 
\sqrt{ 
1 - \frac{\Theta(\vx_i, \vx_j)}{\sqrt{\Theta(\vx_i, \vx_i)\Theta(\vx_j, \vx_j)}}
}.
\end{align}
The RNTK t-SNE pairwise conditional probabilities for two data points $\vx_i$ and $\vx_j$ can then be defined as:
\begin{align}
    p_{j|i} = \frac{\exp\left(\tfrac{-d(\vx_i, \vx_j)^2}{2\sigma_i^2}\right)}{\sum_{k\neq i}\exp\left(\tfrac{-d(\vx_i, \vx_k)^2}{2\sigma_i^2}\right)},
\end{align}
which differs from classical t-SNE simply by replacing its Euclidean distance with the novel RNTK-based distance. Afterwards, the standard t-SNE method of finding the Kullback-Leibler divergence between $p$ and $q$ proceeds as usual to produce the low-dimensional MASK embeddings.

\section{Experiments}

\begin{figure}[t!]
    \centering
    \includegraphics[width=0.8\linewidth]{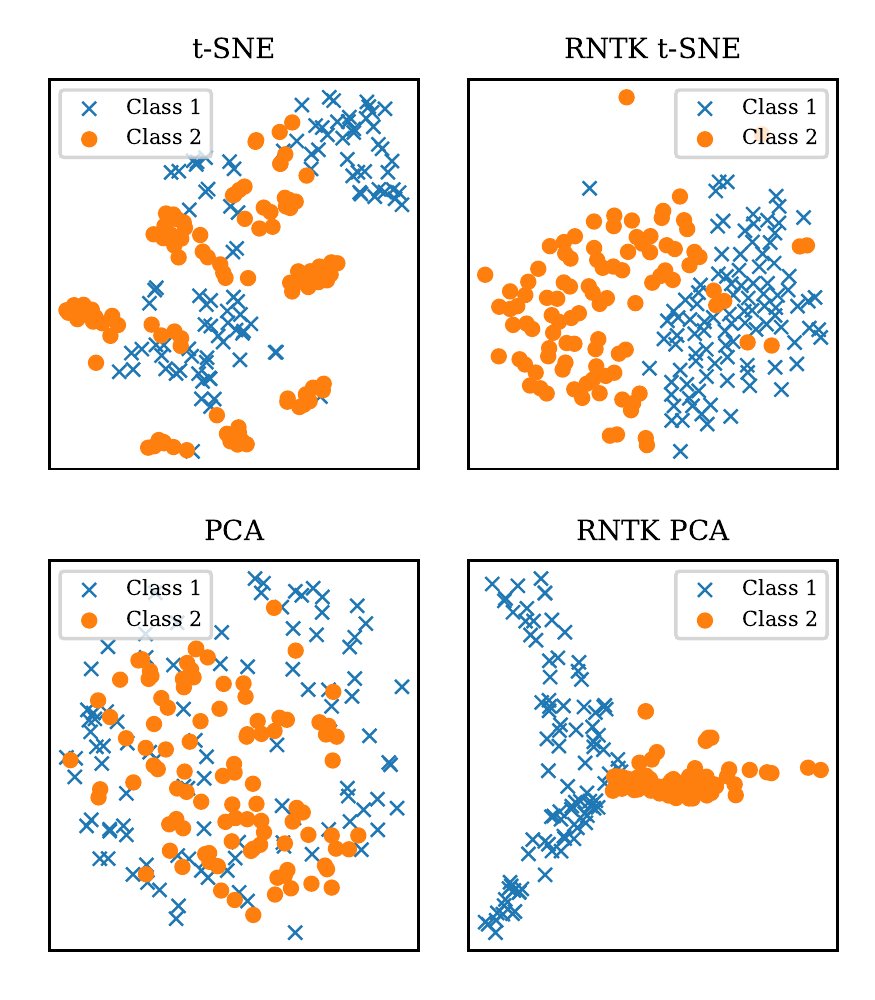}
    \vspace*{-3mm}
    \caption{Demonstration of MASK dimensionality reduction to separate variable-length time series data. \emph{Top left:} t-SNE on original sequences padded with zeros fails to separate the data. \emph{Top right:} RNTK t-SNE successfully separates the data into distinct clusters corresponding to the different classes of time series. \emph{Bottom left:} PCA on original sequences padded with zeros fails to separate the data. \emph{Bottom right:} RNTK PCA successfully separates the data into distinct clusters corresponding to the different classes of time series.}
    \label{fig:tsne_pca_comp}
\end{figure}

We now experimentally demonstrate that two specific examples of MASK, namely RNTK PCA and RNTK t-SNE, produce more distinct patterns on several synthetic datasets compared to PCA and t-SNE with classic kernels.\footnote{Our code can be found at \url{https://github.com/dlej/MASK}.}


\subsection{Dataset}
Many real-world time series data can be modeled by linear dynamic systems. Therefore, for our experiments, we use synthetic time series data generated from
the following family of differential equations:
\begin{align}
 t^{2} \frac{\partial^{2}x(t)}{\partial t^{2}} -  a_{1}  \frac{\partial x(t)}{\partial t} -
a_{0} t x(t) = 0.\label{eq:dyanmics}
\end{align}
To sample time series data of different classes, we draw $a_0 \sim \mathcal{N}(0, 0.0025)$ and $a_1 \sim \mathcal{N}(-4, 4)$ for class 1 and $a_0 \sim \mathcal{N}(0, 0.01)$ and $a_1 \sim \mathcal{N}(-20, 1)$ and for class 2. We sample 100 sequences of length 30 for each class and then randomly drop elements with probability 0.2 in each sequence to shorten the sequences, resulting in sequences with varying lengths. 

We emphasize that we choose the system in Eq.~\ref{eq:dyanmics} only for demonstrative purpose and that MASK works generally on other types of time series data.

\subsection{Experiment Setup}
We compute RNTK using the default set of parameters:
$\sigma_w = 2$, $\sigma_u = 0.316$, and $\sigma_b = 10^{-5}$.
To compare against fixed-length methods, i.e., PCA and t-SNE with classic kernels, we pad the end of the sequences with zeros until they are of the same length. We used the Scikit-learn \cite{pedregosa2011scikit} implementations of PCA and t-SNE with default parameters, although the results were similar even for different values of the method parameters.

\subsection{Results}
Fig.~\ref{fig:tsne_pca_comp} visualizes the dimensionality reduction results comparing PCA and t-SNE with their RNTK replaced counterparts. The color and shapes of points represent different classes. We clearly observe obvious separation of classes in both RNTK PCA and RNTK t-SNE whereas classic PCA and t-SNE fails to separate the data. 
These results validate the proposed MASK's capability in extracting useful patterns from the synthetic data, demonstrating its potential as a practical kernel dimensionality reduction technique for time series data of varying lengths.


\section{Conclusions}

We have demonstrated that, through MASK, we can extend existing, well-proven dimensionality reduction techniques to variable-length data using the RNTK. 
However, dimensionality reduction is not the only method that can be extended in this manner; any kernel- or distance-based method can extended using MASK, for example, clustering. 
Our promising preliminary experimental results on synthetic data validate the utility of MASK. Future avenues of research include evaluating MASK on other modalities of time series data, such as text, audio, speech, and natural language and extending the current technique to other kernel-based methods beyond PCA and t-SNE.
MASK will revolutionize the way we think about dimensionality reduction of different length data sequences and affords us with tremendous future directions for real-world data.

\bibliographystyle{IEEEbib}
\bibliography{refs}

\end{document}